\documentclass[preprint]{aastex}

\shorttitle{Close binary stars.~X}
\shortauthors{Rucinski \& et al.}

\begin{document}

\title{Radial Velocity Studies of Close Binary 
Stars.~X\footnote{Based on the data obtained at the David Dunlap 
Observatory, University of Toronto.}}

\author{Slavek M. Rucinski}
\affil{David Dunlap Observatory, University of Toronto \\
P.O.~Box 360, Richmond Hill, Ontario, Canada L4C~4Y6}
\email{rucinski@astro.utoronto.ca}
\author{Wojtek Pych}
\affil{David Dunlap Observatory, University of Toronto \\ 
and Copernicus Astronomical Center,
Bartycka 18, 00--716 Warszawa, Poland}
\email{pych@camk.edu.pl}
\author{Waldemar Og{\l}oza}
\affil{Mt. Suhora Observatory of the Pedagogical University\\
ul.~Podchora\.{z}ych 2, 30-084 Cracow, Poland}
\email{ogloza@ap.krakow.pl}
\author{Heide DeBond, J. R. Thomson, Stefan W. Mochnacki,
Christopher C.\ Capobianco\altaffilmark{2},\\ George Conidis}
\affil{David Dunlap Observatory, University of Toronto \\
P.O.~Box 360, Richmond Hill, Ontario, Canada L4C~4Y6}
\email{(debond,jthomson,mochnacki,conidis)@astro.utoronto.ca,
ccapo@ap.smu.ca}
\author{P. Rogoziecki}
\affil{Adam Mickiewicz University Observatory, S{\l}oneczna 36, 
60--286 Pozna\'{n}, Poland}
\email{progoz@moon.astro.amu.edu.pl}

\altaffiltext{2}
{Current address: Department of Astronomy and Physics,
Saint Mary's University, Halifax, Nova Scotia, Canada B3H~3C3}

\begin{abstract}
Radial-velocity measurements and sine-curve fits to the orbital 
velocity variations are presented for the ninth set of ten close binary 
systems: V395~And, HS~Aqr, V449~Aur, FP~Boo, SW~Lac, KS~Peg, IW~Per,
V592~Per, TU~UMi, FO~Vir. The first three are very close, 
possibly detached, early-type binaries and all three 
require further investigation. Particularly interesting is
V395~And whose spectral type is as early as B7/8 for a 0.685 day
orbit binary.
KS~Peg and IW~Per are single-line binaries, with the former probably
hosting a very low mass star. 
We have detected a low-mass secondary in an important 
semi-detached system FO~Vir at $q=0.125 \pm 0.005$. 
The contact binary FP~Boo
is also a very small mass-ratio system, $q=0.106 \pm 0.005$.
The other contact binaries in this group are V592~Per, TU~UMi 
and the well known SW~Lac. V592~Per and TU~UMi have bright
tertiary companions; for these binaries, and for V395~And,
we used a novel technique of the broadening functions 
arranged into a 2-dimensional image in phase. The case of TU~UMi
turned out intractable even using this approach and we have 
not been able to derive a firm radial velocity orbit for this 
binary. Three systems of this group were observed 
spectroscopically before: HS~Aqr, SW~Lac, KS~Peg. 
\end{abstract}

\keywords{ stars: close binaries - stars: eclipsing binaries -- 
stars: variable stars}

\section{INTRODUCTION}
\label{sec1}

This paper is a continuation in a series of papers (Papers
I -- VI and VIII -- IX) of
radial-velocity studies of close binary stars 
\citep{ddo1,ddo2,ddo3,ddo4,ddo5,ddo6,ddo8,ddo9} and presents
data for the ninth group of ten close binary stars observed
at the David Dunlap Observatory. For technical details and 
conventions, and for preliminary estimates of uncertainties,
see the interim summary paper \citet[Paper VII]{ddo7}.
Selection of the targets is quasi-random: At a given time, 
we observe a few dozen close binary systems with periods
shorter than one day, brighter than 10 -- 11 magnitude and with
declinations $>-20^\circ$; we
publish the results in groups of ten systems as soon as reasonable
orbital elements are obtained from measurements evenly distributed 
in orbital phases. 

Whenever possible, we estimate spectral types of the program stars
using our classification spectra. These are compared with the
mean $(B-V)$ color indexes taken from the {\it Tycho-2\/}
catalog \citep{Tycho2}
and the photometric estimates of the spectral types
using the relations published by \citet{Bessell1979}.

The observations reported in this paper have been
collected between June 1997 and November 2004. 
The ranges of dates for individual systems can be found in 
Table~\ref{tab1}. 
All systems discussed in this paper, except three: 
HS~Aqr, SW~Lac, and KS~Peg, have been observed by us
for radial-velocity variations for the first time.
We have derived the radial velocities in the same way as described
in previous papers. See Paper~VII for a discussion of the
broadening-function approach used in the derivation of
the radial-velocity orbit parameters:
the amplitudes, $K_i$, the center-of-mass
velocity, $V_0$, and the time-of-primary-eclipse epoch, $T_0$.

The novelty of this paper is the treatment of binaries with
very weak signatures of binarity in their broadening functions,
resulting either from exceptionally weak Mg~I $\lambda 5184$  
triplet lines (V395~And, Section~\ref{V395And}) or from 
the presence of a bright third star in the system 
(TU~UMi, Section~\ref{TUUMi}) which limits the dynamic range in 
the broadening function. We utilize here
an approach borrowed from image processing techniques in that
we represent the broadening function information in the phase
domain as a two-dimensional image where smoothing in the
phase coordinate restores the temporal correlation
between successive broadening functions.

This paper is structured in a way similar to that of previous
papers, in that most of the data for the observed binaries are in
two tables consisting of the radial-velocity measurements in 
Table~\ref{tab1} and their preliminary sine-curve solutions 
in Table~\ref{tab2}. Table~\ref{tab1} contains data for
nine stars as we have not been able to derive individual
radial velocities for the difficult case of TU~UMi.

The data in Table~\ref{tab2} 
are organized in the same manner as in previous papers. 
In addition to the parameters of spectroscopic orbits,
the table provides information about the relation between
the spectroscopically observed epoch of the primary-eclipse T$_0$
and the recent photometric determinations in the form of the $O-C$
deviations for the number of elapsed periods $E$. It also contains
our new spectral classifications of the program objects.
Section~\ref{sec2} of the paper contains brief summaries of
previous studies for individual systems and comments
on the new data. Figure~\ref{fig1} and Figure~\ref{fig4}
show the 2-D broadening function images for V395~And and TU~UMi
while Figure~\ref{fig2} and Figure~\ref{fig3} show the radial 
velocity data and solutions. Figure~\ref{fig5} 
shows the BF's for all systems; the functions have been selected 
from among the best defined ones, usually around the orbital phase of
0.25 using the photometric system of phases counted from 
the deeper eclipse.

\section{RESULTS FOR INDIVIDUAL SYSTEMS}
\label{sec2}

\subsection{V395 And}
\label{V395And}

Photometric variability of the star was discovered by the Hipparcos
satellite \citep{hip}; the star was assigned RRc type
in the 74th Special Name-list of Variable Stars 
\citep{namelist74} on the basis of the Hipparcos data.
\citet{duerbeck1997} suggested that V395~And is a contact 
eclipsing system on the basis of the period-color relation.
Light elements were calculated by \citet{Prib2003}, 
HJD=2448500.4372 +E 0.684656 using the Hipparcos photometric
data and the doubled value of the original period.

V395 And was listed in the \citet{duflot1995}
with the spectral type of this object estimated as B8 
(this type is also quoted in the Washington Double Star 
Catalog\footnote{The Washington Double Star Catalog is
available at the Internet site: http://ad.usno.navy.mil/wds/.}, 
from now on called WDS).
and radial velocity of $-8$ km~s$^{-1}$; in the HD catalog the
spectral type is A0. Our
spectra of the magnesium triplet in V395~And showed an
exceptionally weak, poorly defined, strongly
broadened feature with depth of only 2 percent of 
continuum with no discernible doubling. 
While we expected that the feature will be weak in a nominally
A0 type star (Simbad), the particular weakness of the line was
surprising.

The star is relatively bright, $V_{max}=7.55$. It is
a member of a visual binary (WDS~23445+4623).
The companion at a separation of 5.7 arcsec is 
2.2 mag fainter than V395~And. It is visible 
at the position angle $104^\circ$ and thus is
projected into our East--West spectrograph slit. 
Its presence may have produced weakening of the 
spectral lines by probably less than 15\%.

Original estimates based on the very shallow and poorly defined
broadening functions (BF's) suggested that the star will be entirely 
intractable and may have to be abandoned. Our standard technique 
of measuring radial velocities requires that each BF, for each
spectrum, gives a peak or a pair of peaks for 
radial velocity centroid determination;
such well defined peaks simply did not exist for V395~And.
However, a new approach, partly based on ideas used in image
processing worked relatively well: 
 The BF's were ordered in as columns in a two-dimensional image,
with Y-coordinate being the heliocentric radial velocity.
The X-coordinate, the phase, is obviously unevenly filled
and the data can be interpolated into a regular grid of
phases. At this point,
we utilize the fact that the BF's are correlated in the temporal
(phase) domain so we can use smoothing. Thus, horizontal cuts 
through the BF-image were convolved with a Gaussian kernel
with the full width at half maximum (FWHM) of 0.025 in phase,  
and then re-binned into a new uniform phase system with
40 phase points. The ``broadening function image'' 
is shown in Figure~\ref{fig1} while a section at phase
0.75 is shown in Figure~\ref{fig5}, among the broadening
functions for other stars of this group.

Even after processing of all available 103 spectra of V395~And, 
as described above, only some of the smoothed BF's were measurable
for individual radial velocities. They are listed in
Table~\ref{tab1}
as for other stars, but they should be used with considerable
caution. The smoothed broadening functions were originally 
found at 0.025 intervals in phase and heliocentric times
were calculated for them using the $T_0$ prediction given above. 
Some of the smoothed BF's were still too poor and have been
eliminated from the listing. The final solution is based on
the data as listed in Table~\ref{tab1} and required a  
new value of the initial epoch $T_0$, as given in 
Table~\ref{tab2}. 

The star appears to be a very close pair with
almost identical components, $q=0.88 \pm 0.03$. The component
potentially eclipsed at the primary minimum 
(corresponding to $T_0$) is less massive; 
its peak in the broadening function is slightly sharper
and taller indicating higher surface temperature and lower
rotation rate.  The very small amplitude of light
variations, $\Delta V \simeq 0.04$, suggests that
the eclipses will not be actually observed and that 
the binary is an ``ellipsoidal variable'' showing only
proximity effects.
Our classification spectra clearly indicated a spectral type
as early as B7 which is difficult to reconcile with
the close orbit with the period of 0.685 days. The two spectra 
that we obtained were not identical and showed some
peculiar variability and emission at He~I $\lambda 4009/4027$.
No spectral estimate of $\log g$ was possible due to
peculiarity of the spectra.
Stars of the spectral type of B7 or B8 have masses 
around 2.8 to 3.0 $M_\odot$ and radii around 2.2 to 2.4 $R_\odot$. 
If the orbit has $i \simeq 40^\circ$ then the current data
would imply $A \simeq 6.0 \,R_\odot$ which is just about possible
for a contact system.

\subsection{HS Aqr}

HS~Aqr is a detached or semi-detached binary. Spectroscopic 
studies based on the high quality echelle spectra were presented by
\citet{popper1996,popper1997,popper2000}. This was followed
by an extensive discussion of the photometric and spectroscopic 
data by \citet{clausen2001}. The above sources provide the
most extensive references to properties of the system.
The value of $T_0$ was taken from \citet{Kreiner2004}:
$T_0 = 2,452,500.6960 + E \times 0.710188$. 

Our observations of HS~Aqr give a good solution of
the radial velocity orbit (Figure~\ref{fig2}),
but perhaps not of such a high 
quality as that of Popper which was 
based on the echelle spectrograph spectra.
However, since the main uncertainties in this field are not
in random errors, but in systematic differences in measuring
techniques, we give our solution here for future reference.

The broadening function at phase close to 0.25 (Figure~\ref{fig5})
shows two well defined peaks for detached components, with
that for the secondary components a bit broader indicating that
this component is larger, if this star is in rotation-revolution 
synchronism. This agrees with the supposition \citep{clausen2001}
that the secondary component is close to or fills its Roche lobe. 

Our new spectral type estimate of F6V is slightly earlier than previously
discussed (F8V and G8-9), in a better agreement with the
Tycho-2 \citep{Tycho2} color index $B-V=0.52$.
The binary is in a visual system (CCDM J20409-0036AB, ADS~14147AB), 
but the companion at 1.6 arcsec separation is some 4 magnitudes
fainter than HS~Aqr thus of no consequence for the
radial velocity orbit. The recent minimum prediction 
by \citet{Kreiner2004} perfectly agrees with our determination
of the initial epoch $T_0$.

The star is bright, $V_{max}=9.0$, but was not included in the 
Hipparcos database, so its parallax is currently unknown.

\subsection{V449 Aur}

V449~Aur was discovered photometrically by the Hipparcos mission.
The primary eclipse prediction was the HIP one:
$T_0 = 2,448,500.2670 + E \times 0.703648$. 
The mean $uvby$ data of \citet{Jordi1996}, 
$V=7.455$ and  $b-y=0.086$,
suggest an early-type no later than A5/6V, while our spectral 
classification is A2IV and, through a comparison with
the color index, implies some reddening. 

The binary type of the system is not clear at this moment:
The broadening functions are partially blended, as in a contact 
binary, and somewhat difficult to measure for centroid 
determinations (see Figure~\ref{fig5}).
On the basis of its very shallow Hipparcos light curve, but with
rather well-defined eclipses, the binary was considered by
\citet{ruc02} to be a detached one (EA); the broadening functions
would rather suggest a contact binary. However,
the large widths of the peaks in the broadening functions and
the resulting blending may result from rotational velocities
higher than the synchronous ones. The mass ratio is not 
far from unity, $q=0.730 \pm 0.017$, which is frequently
encountered among early-type very close binaries.   
The scatter of the radial velocity observations of the 
slightly more massive component, which is 
eclipsed in the primary minimum, appears to be genuinely
larger at phases around 0.2 -- 0.4
(Figure~\ref{fig2}); we have no explanation for it.

\subsection{FP Boo}

FP Boo is another case of the Hipparcos photometric discovery.
The Hipparcos period and $T_0$ serve the current observations
relatively well: $T_0 = 2,448,500.478 + E \times 0.640487$.

 The previous spectral type given by  SIMBAD
and in the Hipparcos database, A5, does not agree with our 
estimate of F0V; the latter would imply $B-V=0.28$ whereas
the Tycho-2 mean color index is even redder, $B-V=0.35$, 
corresponding to F2V.

The Hipparcos photometry was analyzed by \citet{Selam2004}
using a simplified approach. The mass ratio was estimated 
at $q_{ph}=0.1$, which is in a surprisingly good agreement with the
spectroscopic result, $q_{sp} = 0.106 \pm 0.005$, 
taking into account that the light curve is shallow and 
does not show any obvious total eclipses. (This case is -- in
our view -- an exception, a fortuitous coincidence.
Normally, we do not trust the so-called photometric 
mass ratios because so many stars
in our RV program show $q(sp) \ne q(ph)$).
FP~Boo was somewhat faint ($V_{max} = 10.07$) for Hipparcos photometry, 
so the light curve could be easily improved. Since the velocity
amplitudes are large and thus imply the orbital inclination close to 
90 degrees, it is likely that more precise photometry
will show total eclipses.  In its small mass ratio 
and the spectral type of F0/2V, the system is somewhat 
similar to the well-known contact binary AW~UMa.

The radial velocity orbit is shown in Figure~\ref{fig2} while
a broadening function close to phase 0.25 is shown in
Figure~\ref{fig5}.

\subsection{SW Lac}

SW Lac is one of the most frequently photometrically observed 
contact binaries. It is a bright ($V_{max}=8.66$) binary
with a large amplitude and an unstable
light curve and thus an easy target for investigations
utilizing small telescopes. Several discussions addressed the matter of
evolving surface spots, the most recent being by 
\citet{Albayrak2004} where references to previous work can be found.

SW Lac is a late type contact binary with the spectral type variously
assigned as G3, G5 and K0. Our new classification is G5V which
predicts a color index a bit less red than the observed
$B-V=0.73$; possibly G8V would be a good compromise.
From the moment of minima predictions 
as well as the direct detection of faint 
spectral signatures \citep{HM1998}, the binary appears to 
have at least one -- probably two -- nearby, low mass companions
\citep{Prib1999}. 
The system is known to change the orbital period. 
We used the moment of primary minimum of 
\citet{Kreiner2004}: $HJD = 2,452,500.1435 + E \times 0.3207158$
which ideally predicts our $T_0$.

Our observations form the third currently
available set of radial velocity data (see Figure~\ref{fig2}). 
\citet{ZL1989} published the first radial
velocity orbit, with $q=0.797 \pm 0.010$. A similar, but not
identical result was obtained soon after by 
\citet{Hriv1992} who published only the mass ratio, 
$q=0.73 \pm 0.01$, without more details. 
In spite of the large brightness of the star and excellent 
definition of the individual peaks in the broadening functions,
our mass ratio, $q=0.776 \pm 0.012$, carries
a larger error than for most of our targets, probably because
of the genuine changes in the broadening functions, which 
are caused by elevated stellar surface activity. What is important,
however, that radial velocity semi-amplitudes 
in our orbit (Figure~\ref{fig2}) are larger than
previously observed: Our total mass estimate,
$(M_1 + M_2) \sin^3 i = 2.10 \pm 0.06 \, M_\odot$ 
(the primary 2.18 $M_\odot$, the secondary
0.92 $M_\odot$), is significantly larger than 
for the orbit of \citet{ZL1989},
$(M_1 + M_2) \sin^3 i = 1.74 \, M_\odot$. Since most
of systematic effects result in a reduction of the 
radial velocity amplitudes, our result is more trustworthy
as it is based on the superior technique of the broadening 
functions. Thus, SW~Lac is a relatively massive, but
surprisingly cool system, a situation which was very clearly
seen before in the case of another cool contact binary
AH~Vir \citep{LR1993}. As in the case of AH~Vir, 
the masses are characteristic for F-type stars 
while the combined luminosity is in agreement with small 
star sizes -- implied by the Roche geometry of a very short
period system -- and their low temperature (G5V). 
Indeed, for SW~Lac, the $M_V$ calibration
\citep{RD97} predicts $M_V(cal) \simeq 4.5$ as compared with the 
parallax-based, observed $M_V(obs) = 4.1 \pm 0.2$, so that
the geometry and photometric properties are
those of a late-type, compact contact system, 
only the masses are too large.

We have not noticed in individual broadening functions
any traces of the third component detected
by \citet{HM1998} and apparently producing part of the 
time-of-minima perturbations \citep{Prib1999,Kreiner2004}. 
However, the same spectra, when averaged in the heliocentric system,
clearly show a late type dwarf contributing about 
1~percent to the total light at our 5184 \AA\ 
spectral window (Caroline D'Angelo and Marten van Kerkwijk,
2005, private communication).

\subsection{KS Peg}

The bright star KS Peg ($V_{max}=5.48$, $B-V=+0.01$), also called
75~Peg, has been known as a radial velocity variable for a long time, 
since the work of \citet{Plaskett1920}.  
\citet{HG1985} recognized the peculiar properties of this
binary star, in particular the 
very small mass ratio of $q < 0.1$ making the velocity amplitude
of the primary component small and the secondary
component invisible in their spectra. 
The photometric variability was observed by
\citet{HMG1988} while a good quality light curve was
provided by the Hipparcos \citep{hip} project. The light
curve with amplitude of only 0.1 magnitude shows very wide
minima with unusual, sharply peaked, brief light maxima. 

We observed KS~Peg in our standard way, and -- because the star
is so bright -- with a particular
attention to detection of the secondary component. In our
series of observations, we have already detected several
binaries with $q \simeq 0.1$, with some even below this limit,
such as SX~Crv at $q \simeq 0.07$ \citep{ddo5}, so we
hoped to see a signature of the secondary. 
The spectral type of KS~Peg is A1V; 
we used the star of A2V as a main template, but we
also used another F8V radial velocity standard to possibly
capture the signature of the secondary component of later
spectral type. None of these led to detection of 
any traces of the secondary, even
when applying the technique of the ``BF-image'' smoothing described
above for V395~And (Section~\ref{V395And}) which
permits detection of weak features correlating in the
phase domain. Since the spectra of such a bright star
result in particularly good determinations of the broadening
functions, the non-detection of the secondary is puzzling and indicates
that the secondary may be a very low mass star or even 
a massive planet, just able to 
enforce rapid rotation on the visible component, but too faint
to contribute light to the overall luminosity of the binary.

One of the panels of Figure~\ref{fig5}
shows comparison of broadening functions at both orbital 
elongations (averages of 6 BF's at each elongation). 
Rapid rotation of the primary component and its (small) orbital
displacements are very clearly visible, but there is no sign
of the secondary component. Note that 
systematic trends in the baselines are due to the 
rectification process with inevitably different tracing
of the continuum for the program and the standard template
spectra. These trends may mask a very weak secondary component.

Our initial reference epoch was that of the Hipparcos photometry,
$T_0 = 2,448,500.286 + E \times 0.502103$, 
which predicts our $T_0$ very well.
In fact, with the same period as in the table, even the epoch
of the \citet{HG1985} is well predicted more than 16 thousand 
epochs earlier (the authors in fact 
suggested 0.5021035, which does not tie all observations that
well). The remarkable constancy of the orbital period 
may be one of useful hints in interpretation of KS~Peg. 

KS~Peg is an important object because it is a 
bright star and thus belongs to the complete, 
magnitude-limited Hipparcos sample of short-period binaries 
brighter than 7.5 magnitude \citep{ruc02}. Its parallax
is moderately large, $13.65 \pm 0.75$ milli-arcsec, and the
star is intrinsically bright, $M_V = +1.1 \pm 0.1$.
The absolute magnitude agrees very well with the expected
value for a Main Sequence A1 star, $M_V \simeq +1.0$.  
The large projected rotational velocity of the primary
component, $V_1 \sin i = 240 \pm 15$ km~s$^{-1}$, implies
that the orbital inclination cannot be far from $90^\circ$
degrees. Assuming $M_1 = 2.3 \pm 0.1 \, R_\odot$, 
$R_1 = 2.3 \pm 0.1 \, M_\odot$,
one can achieve a consistent picture, in terms of reproduction
of $K_1$ and $V_1 \sin i$, with a relatively small 
secondary star with $M_2 = 0.18 \pm 0.01 \, M_\odot$ and thus
$q = 0.078 \pm 0.005$, in a binary with the orbital inclination
$i = 80^\circ \pm 3^\circ$.

\subsection{IW Per}

IW~Per is somewhat similar to KS~Peg described above:
It is also bright, $V_{max}=5.76$, small amplitude, 0.04 mag., 
photometric variable. The variability was recognized by
\citet{mor85}, but most of the research concentrated on
on the chemical peculiarities and Am characteristics
\citep{AM95,ade98,PM98}. 

IW~Per was not a lucky star in our observations. A new CCD
system which was used for it failed after some time
and the data could not be related to those taken
before or after as our technique requires 
high consistency in observations of
program and template stars. Later on, it
was realized, that the heavy dewar of the CCD system induced 
flexure in the spectrograph of up to 1/2 of the pixel in the
detector focal plane. We decided to analyze and publish 
the available data for IW~Per when it was
realized that it is a single-line spectroscopic binary,
precluding any in-depth studies of this system. In the
final radial velocity solution, the spectra taken at 
large hour angles have been given half weight to account
for the spectrograph flexure. The period and the initial epoch
for our observations have been taken from the Hipparcos database. 
$T_0=2,448,500.523 + E \times 0.917180$.

In contrast to KS~Peg, IW~Per is not a very low amplitude
radial velocity variable: The half amplitude $K = 98.6$
km~s$^{-1}$ indicates that the mass ratio is
not very small, probably around $q \simeq 0.45 - 0.5$
 and that the spectroscopically invisible 
component is a star with a mass only slightly smaller
than solar. The visible component
rotates with $V_1 \sin i \simeq 90 \pm 10$ km~s$^{-1}$
which is consistent with synchronous rotation
in a moderately wide binary with the orbital 
period of 0.917 days.

IW Per has the best determined parallax among stars of 
th current group, $18.29 \pm 0.81$ mas. The implied luminosity,
$M_V = 2.07 \pm 0.11$, the spectral type A3V and
$B-V = 0.13$ are mutually consistent.

\subsection{V592 Per}

V592 Per is another photometric discovery of Hipparcos. 
It is a relatively bright, $V_{max}=8.22$,  
basically unstudied contact binary.
The moment of primary eclipse from Hipparcos,
$HJD = 2,448,500.243 + E \times 0.715722$ serves
well for prediction of $T_0$. The
star has been known to possess a visual companion at
a separation of only 0.16 arcsec \citep{Heintz1990}. 
Data for WDS~04445+3953 indicate that the position
angle changes rather rapidly so the system is definitely
a physical one.
We re-discovered the companion spectroscopically: In the
broadening functions, its signature is strong, because of
its sharpness, but the companion is fainter
than the binary with $L_3/(L_1+L_2) = 0.60 \pm 0.06$,
or $\Delta m = 0.55$ mag; in the WDS catalog
$\Delta m = 0.85$ mag.

To determine the close binary orbit, we analyzed 
our observations the way as before for triple systems, 
by first fitting
three Gaussians to the whole BF, then subtracting the
third star Gaussian and then measuring the radial velocities
of both binary components from the residual broadening 
functions. In spite of the presence of the visual
companion, the contact binary 
orbit is well defined, with $q = 0.408 \pm 0.008$. The
spectroscopic result is in total disagreement with the mass ratio 
estimated from the light curve of $q_{ph}=0.25$ 
\citep{Selam2004}.
Once again, we stress that -- due to the very low
information content of light curves -- light curve solutions for
partially eclipsing contact binaries have little sense
and they should not be attempted without supporting
spectroscopic data.

Our observations are spread over time in that there is
a gap of two full seasons. 
The radial velocities of the third component were
estimated from all available spectra giving
the average value of $V_3=+29.28 \pm 0.20$ km~s$^{-1}$
for the first season ($2,451,870 - 2,451,957$)
and $V_3 = 27.92 \pm 0.18$ km~s$^{-1}$
for the second season ($2,452,513 - 2,452,517$).
Our systematic uncertainties
are at the level of $1.5 - 2.0$ km~s$^{-1}$, while 
measurements in the presence of the variable binary
``background'' difficult, so we cannot
claim that we see seasonal variation in $V_3$; it is 
basically identical to $V_0$ of the contact binary.

In view of similar luminosities, 
the spectral type F2IV estimated by \citet{Grenier1999} most
probably applies to both stars, the contact binary V592~Per
and the third star in the system. This spectral type 
does not agree with $(B-V) = 0.47$ estimated from 
the Tycho-2 data \citep{Tycho2} which suggests F5--F6.

\subsection{TU UMi}
\label{TUUMi}

TU UMi is one of the Hipparcos photometric discoveries.
A description of new photometric observations conducted
concurrently with the spectroscopic observations 
was given by \citet{PR2004}; the paper contains also references
to work related to identification of the variability and of the
determination of the period at twice the Hipparcos
original period.

The previous photometric paper \citep{PR2004} was
based on a short time interval and could not provide a new value
of the period which was estimated to only 4 significant figures.
However, the Hipparcos period, which we use here, together with
the new photometric epoch, JD(min) = 2,452,725,6262 \citep{PR2004},
counted back to the Hipparcos epoch (additionally
shifted by 1/4 of the period because the star 
was thought to be a pulsating one so the time of its
maximum light was used), JD(min) = 2,448,499.9747, indicate
a good stability of the period.

TU UMi is a case of a contact binary with a bright close 
companion. We were not aware that it had been known as
a visual binary WDS~14557+7618 with separation of 0.2 arcsec 
(this is well below the 1.88m telescope resolution in our
location) when we started the observations.
The spectral lines of the close binary are very broad 
and, in fact, entirely invisible in individual spectra,
being masked by the strong sharp spectrum of the companion. 

With 70 individual spectra rather evenly covering orbital
phases, we attempted to analyze the data using the approach similar
to that described for V395~And (Section~\ref{V395And})
by ordering in phase and smoothing the broadening functions. 
This case is much more difficult than that of V395~And because,
even after averaging, the binary features are too weak 
and too poorly defined for being measured;
most of the dynamic range in the BF is ``wasted'' on the sharp peak
of the tertiary component (Figure~\ref{fig5}). 
We attempted to remove this peak
from the broadening function. First, we followed
our normal routines of approximating the third star feature by
a Gaussian and subtracting it. This did not work as the
profile is not Gaussian at its base where its shape is
particularly important. Then we
attempted to find the shape of the third-star peak
by combining all broadening functions and determining their
lower envelope. This still did not provide a correct
shape of the tertiary peak at the base, since none
of the BF's was really free of the binary signature. 
Thus, we were unable to remove the tertiary
and to determine individual radial velocities of
components for TU~UMi; this star does not have any
data listed in Table~\ref{tab1}.
While this is disappointing, we feel that
it is doubtful if this star will ever permit a detailed 
parameter determination. In this situation, we present 
only very rough estimates from what can be derived from 
just eyeballed fit to the broadening functions in the
2-dimensional representation (Figure~\ref{fig4}). 

In plotting the expected changes in the broadening
functions as a function of phase, we assumed the initial
epoch as observed simultaneously
photometrically by \citet{PR2004}. 
The star appears to be a W-type contact system with a rather
large amplitude of the secondary star variation, perhaps
as large as $K_2 \simeq 220 \pm 20$ km~s$^{-1}$. 
An estimate for the primary is $K_1 \simeq 35 \pm 15$ 
km~s$^{-1}$. These amplitudes are shown on the BF image in
Figure~\ref{fig4}. 
Integrations of the wide binary feature at several phases 
indicate that the binary is in fact only marginally fainter
than the third component, $L_3/(L_1+L_2) = 1.25 \pm 0.15$,
but the sharpness of its peak dominates in the combined
spectrum of the system. The above estimate of the relative
luminosity may carry a large systematic error since it is 
strongly dependent on how the baseline of the broadening function is
drawn; in fact, this luminosity ratio implies 
$\Delta m = -0.24 \pm 0.12$, which poorly agrees with 
the direct magnitude difference given in the WDS catalog 
($\Delta m = -0.47$).

The third component was observed to have the mean velocity of
$-4.16 \pm 0.20$ km~s$^{-1}$ (a single observation $\sigma = 1.7$
km~s$^{-1}$).

\subsection{FO Vir}

FO~Vir is a totally eclipsing, detached or 
semi-detached binary, with the larger component close to filling
the Roche lobe, thus possibly an object in the pre-contact
state. An extensive discussion of its properties, 
following the photometric study of \citet{SF1984} who
established its binary character, was presented by
\citet{Mki1986}. On the basis of photometry and
radial velocities of the dominant A7V component, they were
able to infer many parameters of the system, in particular
they predicted the mass ratio, $q = 0.15 \pm 0.02$, in 
a semi-detached system with the more massive component
filling or almost filling its Roche lobe. Similar 
results were obtained by \citet{Por1987}.

We have been able to detect the secondary component of FO~Vir, 
but with some difficulty, because it is relatively small 
and faint at the
observed mass ratio $q = 0.125 \pm 0.005$ and much later
spectral type than that of the A7V primary, probably early K. 
A somewhat indirect detection of the secondary was
reported by \citet{Shaw1991} who observed the Lyman-alpha
$\lambda1216$ emission moving in anti-phase with the primary
star and thus most likely related to the secondary; their
$q = 0.16 \pm 0.04$ is in agreement with the current 
determination.

We experienced difficulties with measuring the radial velocities
of both components in the first half of the orbit; this could
be related to the previously reported strong and
variable O'Connell effect \citep{Shaw1991}. 
In addition, the case of FO~Vir confronted us with
what may be a limitation of our broadening functions approach:
We observed relatively strong (similar to those of
the secondary component), initially unexplained
broadening-function features moving in phase with the primary. 
We suspect that these ``ghosts'' may appear because of
the low metallicity of the primary component
\citep{Mki1986} and a mismatch between the spectra of
program and template stars; we simply could not
locate a proper, low metallicity template among bright 
standard stars. Many tests and experiments with different
templates give us assurance that the final orbital
elements are secure ones, although certain element of
uncertainty still remains. 

The radial velocity data for FO~Vir may have
to be re-interpreted when a better $T_0$ prediction for the
duration of our observations (1997 -- 2001) becomes
available. With observations spanning four seasons, we could
establish that the period must be shorter than that
used by the Hipparcos project (0.775569) and we used
one which is only by one digit in the last place shorter
than that; however, the period may be -- in fact -- systematically
changing so that a full discussion is clearly in order, especially
in view of a possibly sporadic mass transfer in the system.
Our $T_0$ is shifted by 50 minutes from the time predicted
by the Hipparcos project. A re-discussion of rather few
available moments of minima for 1983 -- 1991
gave $JD(hel) = 2,452,500.1514 + E \times 0.7755678$, which
we took as a guidance for adopting a shorter period, but which
nevertheless resulted in the spectroscopic $T_0$ for our
program too early by 40 minutes. 

FO~Vir is a member of a visual binary WDS~13298+0106, but the
companion is some 6.5 fainter so that its presence was of no
consequence for the spectroscopic observations.

\section{SUMMARY}

This paper presents radial-velocity data and orbital solutions 
for the ninth group of ten close binary systems 
observed at the David Dunlap Observatory. This group contains
very interesting objects: KS~Peg and IW~Per are single-line (SB1)
binaries without any traces of companions; the former may have
a very small-mass secondary ($q < 0.1$).
V395~And, HS~Aqr and V449~Aur are very close, early-type 
binaries; the matter of them being detached systems requires
further work; V395~And with its spectral type of B7/8 is
particularly interesting because of its orbital period of
only 0.685 days. FP~Boo, SW~Lac, V592~Per
and TU~UMi are contact binaries. Two of them, V592~Per and TU~UMi,
have bright tertiary companions, while FP~Boo is a very small
mass-ratio ($q=0.11$) contact binary. The frequently observed
system SW~Lac is more massive than estimated 
before; its broadening functions show the large
surface brightness and rather compact dimensions of its
secondary component, both possibly indicating that it is
really a semi-detached binary and not a contact system. 
Finally, we have been able
to detect the low-mass secondary in FO~Vir ($q = 0.12$);
the system is most probably a semi-detached binary 
before establishing contact.

\acknowledgements

We express our thanks to Greg Stachowski for observations and
Staszek Zola for advice on handling the case of KS~Peg.

Special thanks go to Prof.\ Janusz Kaluzny who, acting as a referee,
made very useful suggestions and comments which improved 
presentation of this paper.

Support from the Natural Sciences and Engineering Council of Canada
to SMR and SWM and from the Polish Science Committee (KBN
grants PO3D~006~22 and P03D~003~24) to WO and RP
is acknowledged with gratitude. 

The data given in this paper were obtained and analyzed
when WP held the NATO Post-doctoral Fellowship administered by
the Natural Sciences and Engineering Council of Canada.

The research made use of the SIMBAD database, operated at the CDS, 
Strasbourg, France and accessible through the Canadian 
Astronomy Data Centre, which is operated by the Herzberg Institute of 
Astrophysics, National Research Council of Canada.
This research made also use of the Washington Double Star (WDS)
Catalog maintained at the U.S. Naval Observatory.

\clearpage

\noindent
Captions to figures:

\bigskip

\figcaption[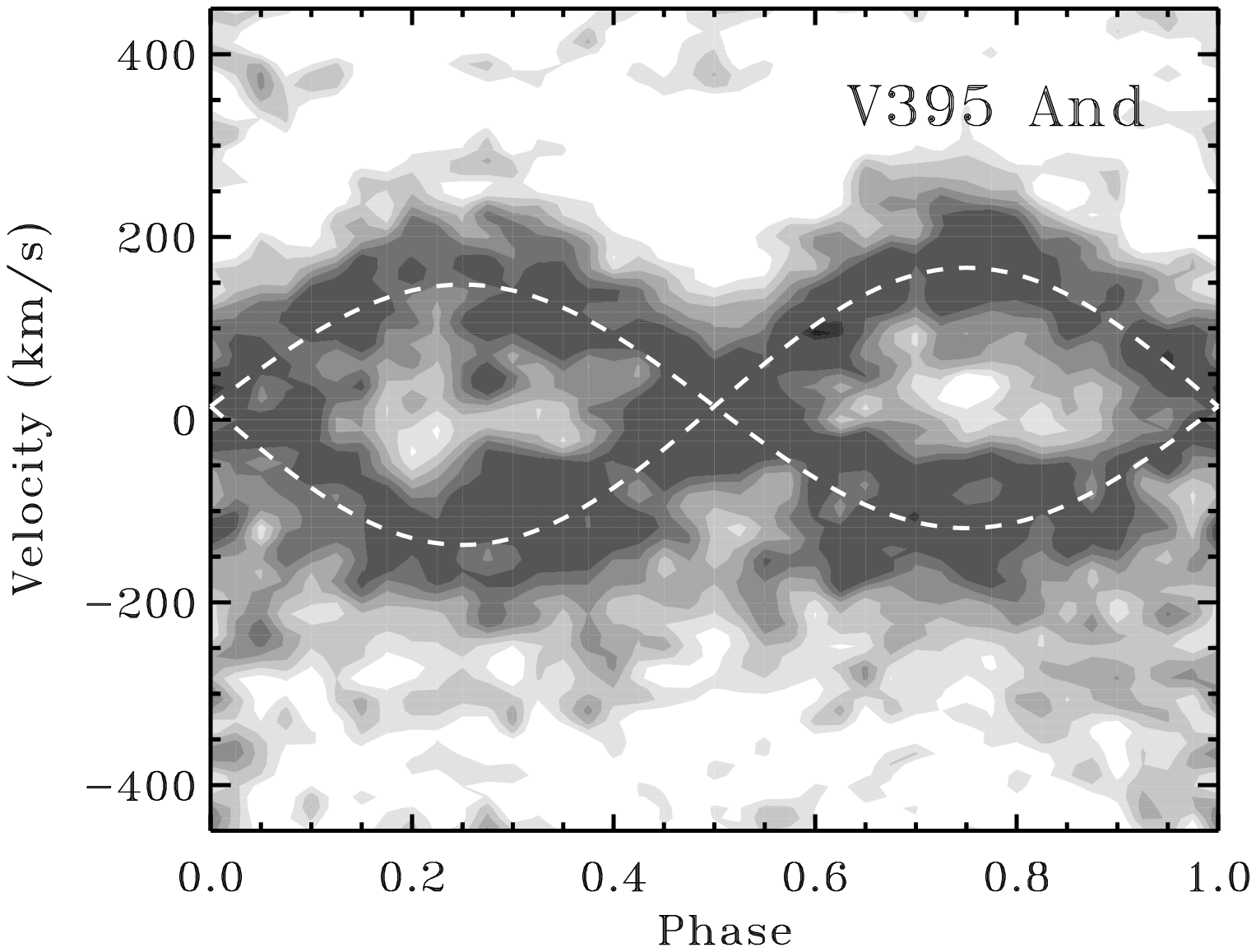] {\label{fig1}
The broadening function ``image'' for V395~And with the
orbital solution superimposed on the contour plot. The image
has been smoothed in the phase domain with a Gaussian of
FWHM = 0.025 in phase. A vertical  
section of the image at phase 0.75 is 
shown in Figure~\ref{fig5}. 
}

\figcaption[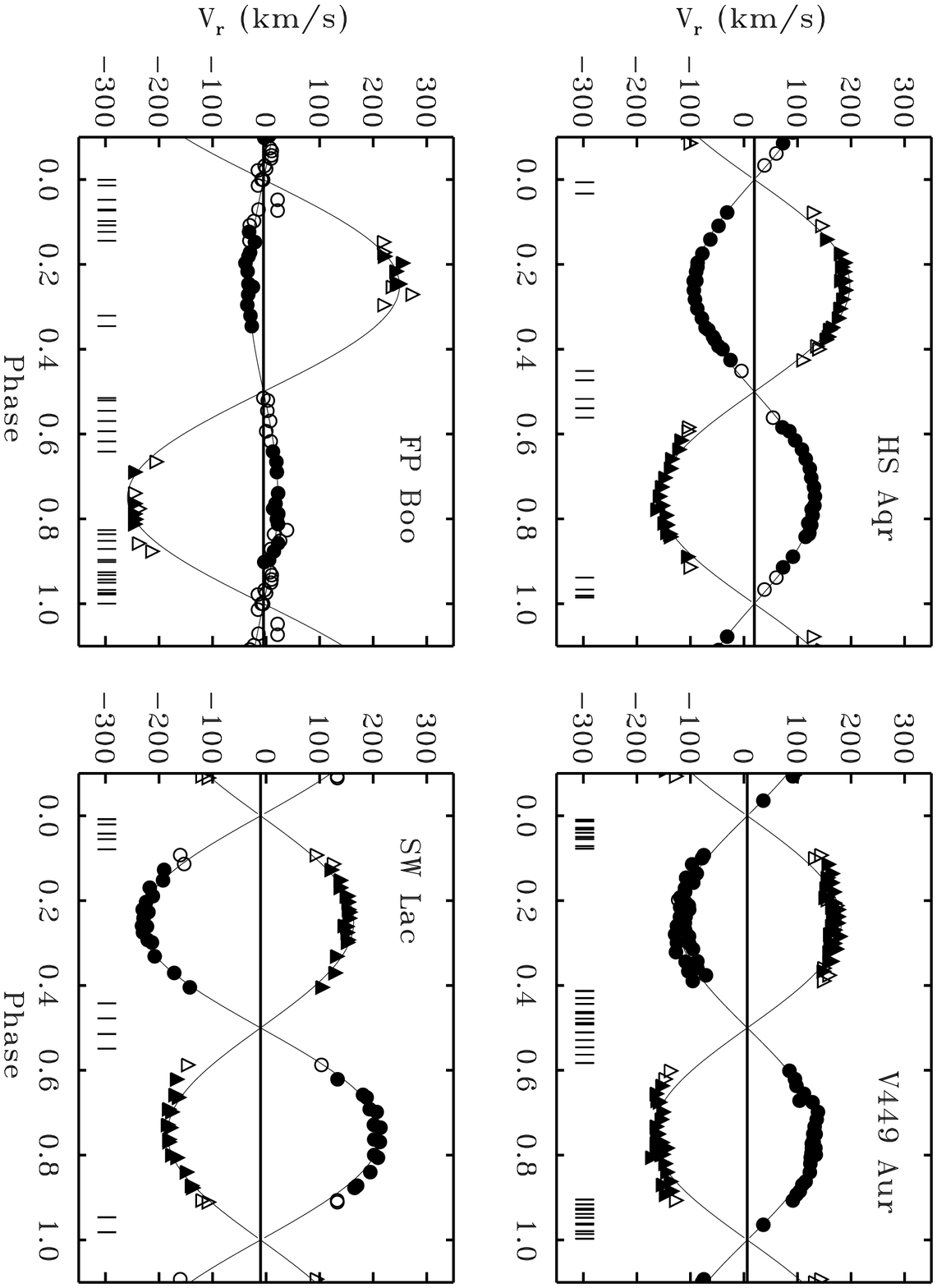] {\label{fig2}
Radial velocities of the systems HS~Aqr, V449~Aur,
FP~Boo and SW~Lac are plotted in individual panels 
versus the orbital phases. 
The lines give the respective circular-orbit (sine-curve) 
fits to the radial velocities. 
HS~Aqr and V449~Aur are detached binaries, while
FP~Boo and SW~Lac are contact binaries.
The circles and triangles in this and the next two figures
correspond to components with velocities $V_1$ and $V_2$,
as listed in Table~\ref{tab1}, respectively. 
The component eclipsed at the minimum corresponding to 
$T_0$ (as given in Table~\ref{tab2}) is the one which shows
negative velocities for the phase interval $0.0 - 0.5$.
The open symbols indicate observations contributing 
half (or less) weight in the orbital solutions. 
Short marks in the lower 
parts of the panels show phases of available 
observations which were not used in the solutions because of the 
blending of lines. All panels have the same vertical 
range, $-350$ to $+350$ km~s$^{-1}$. 
}

\figcaption[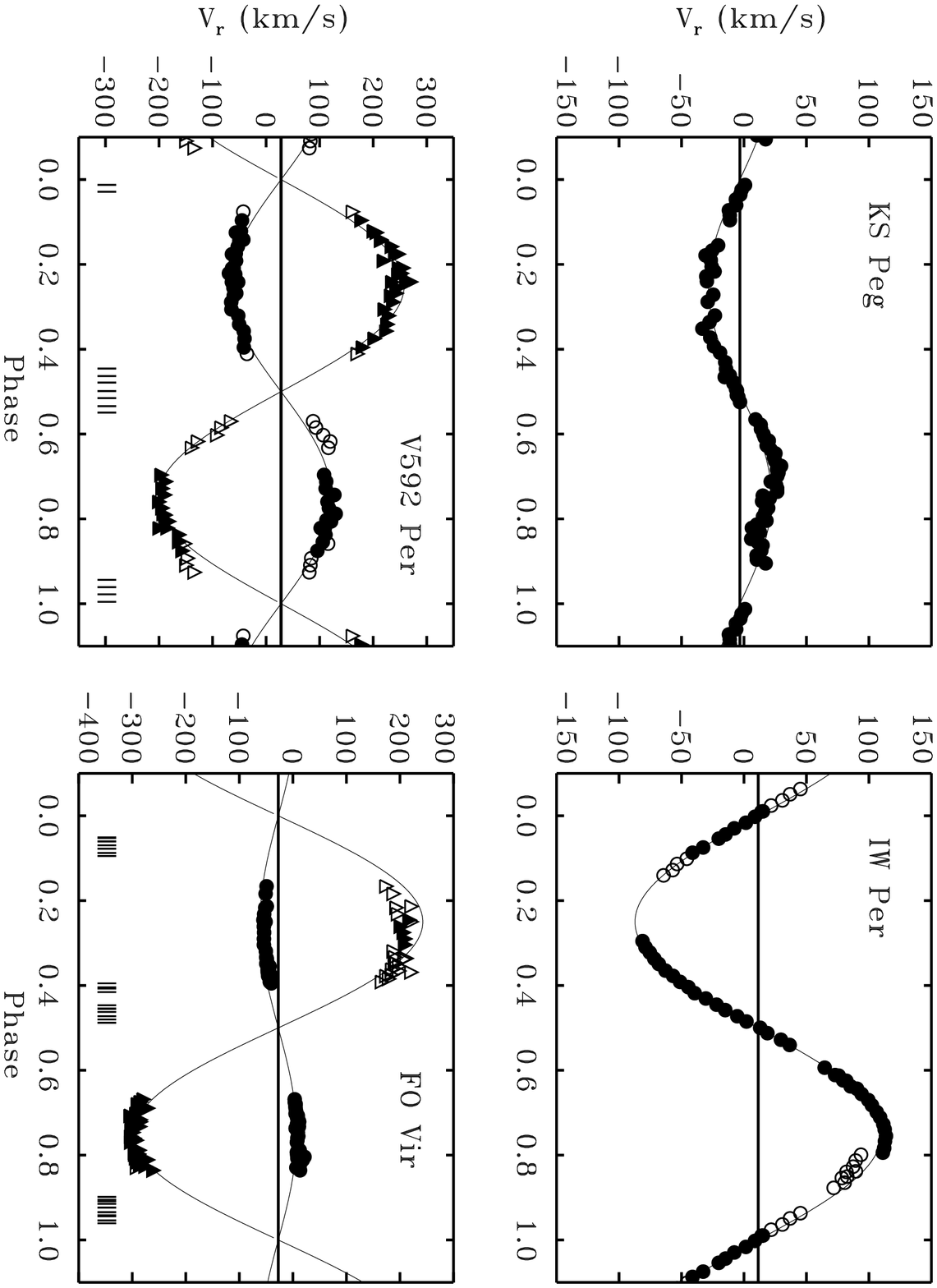] {\label{fig3}
Same as in Figure~\ref{fig2}, but for KS~Peg, IW~Per,
V592~Per and FO~Vir. Note that the vertical scale is
different and much expanded for the single-line
binaries KS~Peg and IW~Per,
while for FO~Vir it is the same for the other systems, but
shifted vertically by $+50$ km~s$^{-1}$. 
}

\figcaption[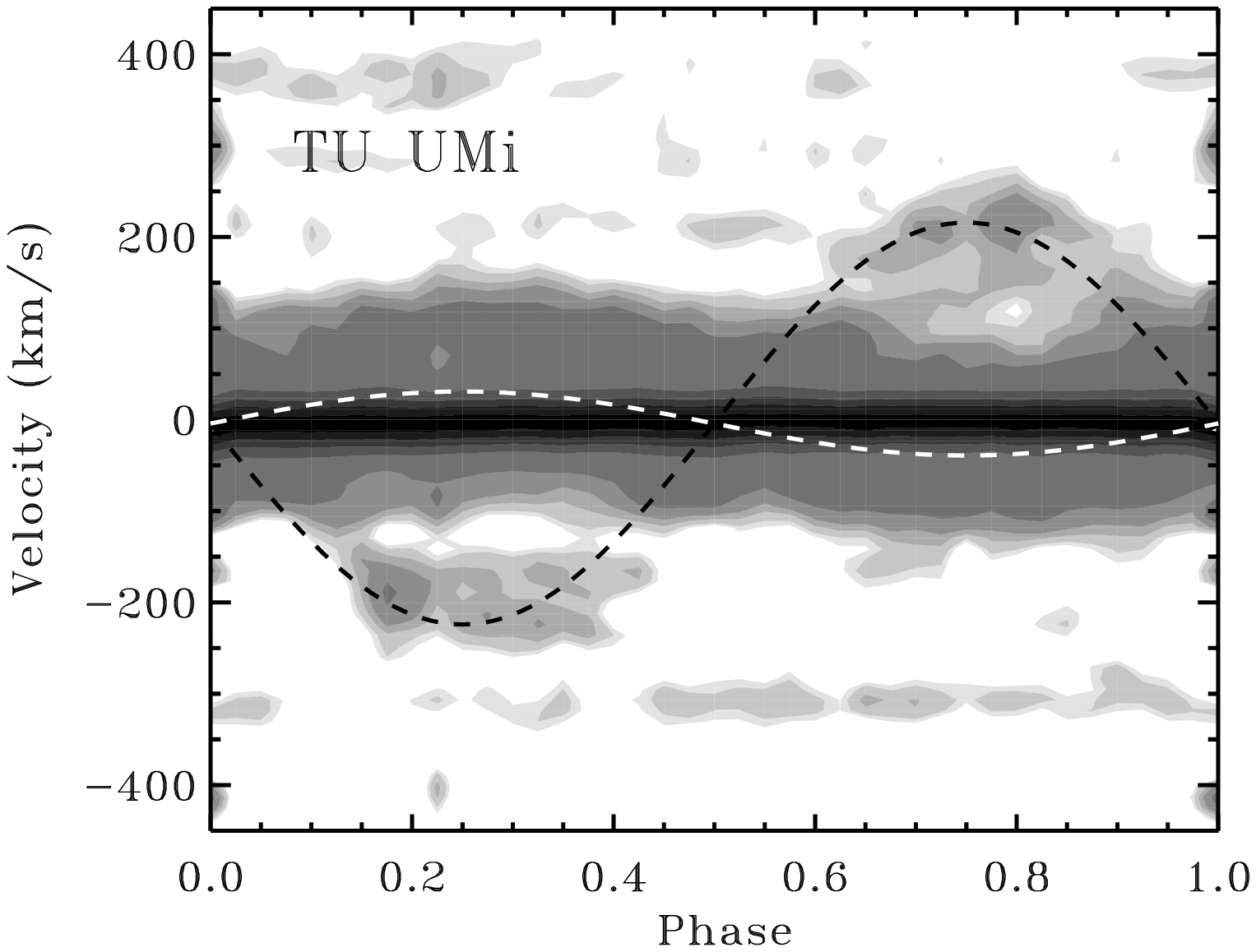] {\label{fig4}
The broadening function ``image'' for TU~UMi with a very
approximate guess at an orbital solution 
superimposed on the contour plot. The image
has been smoothed in the phase domain with a Gaussian of
FWHM = 0.025 in phase. Sections at phases 0.25 and 0.75 are 
shown in Figure~\ref{fig5}. 
}

\figcaption[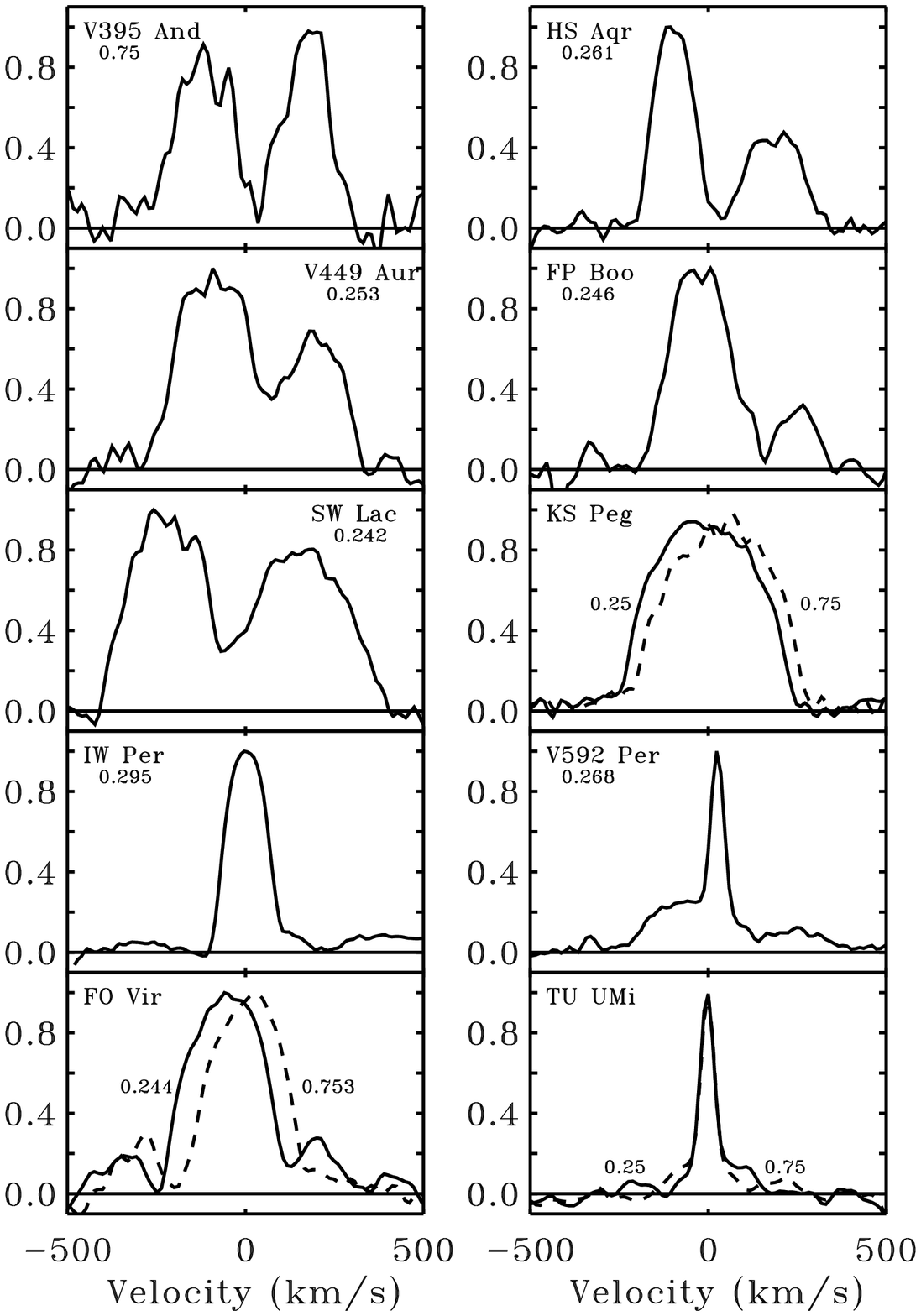] {\label{fig5}
The broadening function for all ten systems of this
group. For most systems broadening functions
at phases close to 0.25 have been selected. The phases
are marked by numbers in individual panels; for
KS~Peg, FO~Vir and TU~UMi the numbers are written 
above locations of the secondary star peaks.  
For V395~And, KS~Peg and TU~UMi, the plots give actually 
sections of the ``BF images'', after Gaussian 
smoothing (FWHM = 0.025) in the phase domain. For  
V395~And the section at phase 0.25 was too poorly defined
so the section at phase 0.75 is shown instead.
Note the high brightness of the secondary in
SW~Lac which directly shows the W-type phenomenon;
note that this could be also interpreted by the
semi-detached configuration of this binary (which
otherwise is considered  as one of the most typical contact
systems). 
The uneven baseline for IW~Per is due to
differences in continuum tracing for the star and the
template. 
As described in the text, FO~Vir showed spurious peaks 
(one is visible here),
which were almost as strong as those for the
secondary component, but followed the motion 
of the primary component; we think it is because of
the mismatch of the template with the low metallicity
spectrum of FO~Vir and because of the continuum height
different for the template and for the star itself. 
}

\clearpage

\begin{deluxetable}{lrrrr}

\tabletypesize{\footnotesize}

\tablewidth{0pt}
\tablenum{1}

\tablecaption{DDO radial velocity observations (the full
table is available only in electronic form)
\label{tab1}}
\tablehead{
\colhead{HJD--2,400,000}    & 
\colhead{~V$_1$} & \colhead{~~W$_1$} &
\colhead{~V$_2$} & \colhead{~~W$_2$} 
}
\startdata
\sidehead{\bf HS Aqr}
2452507.6792  &  122.09  &    1.00  & -143.05  &    1.00 \\
2452507.7194  &   91.43  &    1.00  & -104.07  &    1.00 \\
2452507.7372  &   73.05  &    1.00  &  -99.45  &    0.50 \\
2452507.7543  &   60.43  &    0.50  &    0.00  &    0.00 \\
2452507.7744  &   38.30  &    0.50  &    0.00  &    0.00 \\
2452507.7875  &    0.00  &    0.00  &    0.00  &    0.00 \\
2452509.6398  &   84.74  &    1.00  & -103.00  &    0.50 \\
2452509.6551  &   95.57  &    1.00  & -116.69  &    1.00 \\
2452509.6704  &  108.27  &    1.00  & -120.46  &    1.00 \\
2452509.6863  &  115.38  &    1.00  & -133.96  &    1.00 \\
\enddata

\tablecomments{The table gives the radial velocities $V_i$ and
associated weights $W_i$ for observations described in the paper. 
The first 10 rows of the table for the second
program star, HS~Aqr, are shown; this star is more typical
than the first, V395~And, for which radial velocities
were obtained from smoothed, averaged broadening functions
as described in the text. There are no 
radial velocity data for TU~UMi.
In the table, velocities are expressed in km~s$^{-1}$. 
Observations leading to entirely inseparable 
broadening- and correlation-function peaks 
are given zero weight; these observations
may be eventually used in more extensive modeling of broadening
functions. The radial velocities designated as $V_1$ correspond
to the component which was stronger and easier to measure in
the analysis of the broadening functions; it was not always the
component eclipsed during the primary minimum at the
epoch $T_0$ (see Table~2). The figures should help in
identifying which star is which.}

\end{deluxetable}

\begin{deluxetable}{lcccccccc}

\tabletypesize{\scriptsize}

\pagestyle{empty}
\tablecolumns{9}

\tablewidth{0pt}

\tablenum{2}
\tablecaption{Spectroscopic orbital elements \label{tab2}}
\tablehead{
   \colhead{Name} &                
   \colhead{Type} &                
   \colhead{Other names} &         
   \colhead{$V_0$} &               
   \colhead{K$_1$} &               
   \colhead{$\epsilon_1$} &        
   \colhead{T$_0$ -- 2,400,000} &  
   \colhead{P (days)} &            
   \colhead{$q$}          \\       
   \colhead{}     &                
   \colhead{Sp.~type}    &         
   \colhead{}      &               
   \colhead{} &                    
   \colhead{K$_2$} &               
   \colhead{$\epsilon_2$} &        
   \colhead{(O--C)(d)~[E]} &         
   \colhead{$(M_1+M_2) \sin^3i$} &    
   \colhead{}                         
}
\startdata

V395 And\tablenotemark{a} & EA & HD~222900  & $+14.60$(1.51) & 133.35(2.55) &
     14.20 & 53,278.0000(30)  & 0.684656   & 0.879(25) \\
           & B7/8:            & HIP~117111 &           & 151.72(2.06) &
      9.09 & $+0.0332$~[6978] & 1.647(80)  &           \\[1mm]

HS Aqr     & EA               & HD~197010  & $+19.34$(0.64) & 111.57(0.56) &
      3.36 & 52,507.7981(12)  & 0.710188   & 0.626(7)  \\
           & F6V              & BD$-01^\circ 4025$ &        & 178.24(1.46) &
      9.94 & $+0.0002$~[10]   & 1.795(38)  &           \\[1mm]

V449 Aur   & EW(A)            & HD~41578   & $+6.24$(1.22)  & 128.27(1.45) &
      9.72 & 52,307.0853(24)  & 0.703648   & 0.730(15) \\
           & A2/3V            & HIP~29108  &           & 175.61(2.15) &
     16.76 & $+0.0826$~[5410] & 2.051(73)  &           \\[1mm]

FP Boo     & EW(A)            & BD$+43^\circ 2523$ & $-4.87$(1.02)  & 26.88(0.98) &
      8.08 & 52,388.2227(36)  & 0.6404870  & 0.106(5)  \\
           & F0V              & HIP~76970  &           & 254.04(2.80) &
     14.28 & $-0.0114$~[6070] & 1.475(60)  &           \\[1mm]

SW Lac     & EW(W)            & HD~216598  & $-10.32$(1.17) & 173.91(1.78) &
     10.11 & 52,484.1080(8)   & 0.3207158  & 0.776(12) \\
           & G5V              & HIP~113052 &           & 224.14(1.70) &
     11.54 & $+0.0003$~[-50]  & 2.101(55)  &           \\[1mm]      

KS Peg     & SB1 (EB?)        & HD~222133  & $-3.27$(0.62) & 25.19(0.80) &
     5.25  & 50,735.6405(28)  & 0.502103   &           \\
           & A1V              & HIP~116611 &           &              &
           & $-0.0059$~[16,115] &          &           \\[1mm]

IW Per     & SB1 (EB?)        & HD~21912   & $+11.44$(0.31) & 98.64(0.51) &
     3.46  & 53,031.4192(8)   & 0.91718    &           \\
           & (A3V)            & HIP~16591  &           &              &
           & $+0.0270$~4940]  &            &           \\[1mm]

V592 Per   & EW(A)            & HD~29911   & $+27.92$(1.04) & 93.74(1.25) &
     9.42  & 52,516.8641(20)  & 0.715722   & 0.408(7)  \\
           & (F2IV)           & HIP~22050  &           & 229.97(1.85) &
     15.00 & $-0.0108$~[5612] & 2.521(72)  &           \\[1mm]

TU UMi\tablenotemark{b} & EW(W)   & BD$+76^\circ 544$ & $-4.16$(0.20) & 35(15) &
    30     & 52,725.6262      & 0.377088   & 0.16(7)   \\
           & (F2)             & HIP~73047  &           & 220(20) &
    40     & 0~[0]            & 0.65(27)   &           \\[1mm]   

FO Vir     & EA/EB            & HD~117362  & $-27.26$(1.45) & 33.84(1.02) &
     6.75  & 51,660.1839(25)  & 0.775568   & 0.125(5)  \\
           & (A7V)            & HIP~65839  &           & 269.91(2.95) &
    15.87  & $-0.0342$~[4074] & 2.257(89)  &           \\[1mm]

\enddata
\tablenotetext{a}{V395 And: The solution is based on
103 data points arranged in phase using the preliminary
$T_0$ and period, as given in the text, and then 
with the broadening functions smoothed in the phase domain.
Such averaged data points are given in Table~1.}

\tablenotetext{b}{TU UMi: The data do not permit a
solution; only very rough estimates are given. The value
of $V_0$ is for the visual companion, while $T_0$ is
from the photometric solution based on data
obtained simultaneously with the spectroscopic 
program \citep{PR2004}. Because of the low contribution
of the close binary to the total light, the (literature) 
spectral type is that of the visual companion.}

\tablecomments{The spectral types given in column two
relate to the combined spectral type of all components 
in a system; they are given in parentheses if
taken from the literature, otherwise are new,
The convention of naming the
binary components in the table is that the more massive
star is marked by the subscript ``1'', so that 
the mass ratio is defined to be always $q \le 1$. The figures
should help identify which component is eclipsed at the primary
minimum. The standard errors of the circular 
solutions in the table are expressed in units of last decimal places 
quoted; they are given in parentheses after each value. 
The center-of-mass velocities ($V_0$), the 
velocity amplitudes (K$_i$) and the standard unit-weight 
errors of the solutions ($\epsilon$) are all expressed 
in km~s$^{-1}$. The spectroscopically determined 
moments of primary minima are given by $T_0$; the corresponding
$(O-C)$ deviations (in days) have been calculated from the 
available prediction on $T_0$, as given in the text,
using the assumed periods and the number of epochs given by [E].
The values of $(M_1+M_2)\sin^3i$ are in the solar mass units.
}

\end{deluxetable}

\end{document}